\documentclass[twoside]{article}
\usepackage{fleqn,espcrc2}

\usepackage{epsfig}
\usepackage[figuresright]{rotating}

\newcommand{\AmS}{{\protect\the\textfont2
  A\kern-.1667em\lower.5ex\hbox{M}\kern-.125emS}}

\title{Charmed baryons from lattice QCD}

\author{R.M. Woloshyn\address{TRIUMF, 4004 Wesbrook Mall, Vancouver,
                              BC V6T 2A3, Canada }%
        }
       
\begin{document}

\begin{abstract}
The results of a recent quenched lattice QCD simulation
for charmed baryons are presented. In contrast to
a previous calculation, it is found that hyperfine
splittings are in agreement with quark model expectations
and comparable to experimental values. Preliminary
calculations using lattice NRQCD yield results which are
compatible with those obtained using a Dirac-Wilson action
of the D234 type for the charm quark. 
\vspace{1pc}
\end{abstract}

\begin{flushright}  
\setlength{\baselineskip}{2.6ex}
TRI--PP--00-58\\
Oct 2000\\
\end{flushright}

\vspace{0.8cm}

\begin{center}
{\huge  \bf Charmed baryons from lattice QCD}\\
\vspace{0.8cm}
{R.M. Woloshyn}\\ 
\vspace{0.3cm}
{\small\em TRIUMF, 4004 Wesbrook Mall, Vancouver, BC V6T 2A3, Canada\\}
\end{center}
\vspace{0.8cm}
\noindent
Presented at the IVth International Conference on Hyperons, Charm and Beauty
Hadrons, Valencia, Spain,  27th to 30th, June, 2000.

\maketitle

There have been many applications of lattice QCD to heavy-flavoured hadrons
but most of the work has been restricted to the meson sector. 
The most complete study of heavy baryons in a relativistic framework 
is the one done in the UKQCD
collaboration\cite{ukqcd1} using an \( O(a) \)
improved fermion action in quenched approximation. In that work
the overall features of the baryon spectrum were found to be in
agreement with experiment but 
hyperfine splittings in both charm and bottom baryons were found to be 
considerably
smaller than those predicted by phenomenological models\cite{martin} and
observed experimentally. In fact, the central values for hyperfine splittings
found in \cite{ukqcd1} were negative. Such a result is difficult to understand
and would pose a severe problem for lattice QCD.
Therefore further investigation seems justified.

Due to limitations in computing resources it was not possible in this work
to use the same lattice spacing and volume as used in the UKQCD calculation.
Rather, a special strategy had to be adopted.
We work on a more coarse lattice (\( \sim  \)0.2fm) with a highly improved
action. Past experience has shown that results of reasonable accuracy may be
obtained with such lattices\cite{coarse}. In order to check the calculation, the spectrum
of baryons in the light quark (u,d,s) sector was calculated at the same time.
As well, meson masses for both heavy and light quarks were calculated. 
The results of all these calculations (see \cite{cbary} for details) are in 
reasonable agreement with experimental values
and with the results obtained at small lattice spacing\cite{cppacs,ukqcd2}.

The explicit form of the gluon and fermion action may be found in \cite{cbary}.
The fermion action is of the D234 type\cite{alford} with tadpole improvement
implemented using the Landau link. The lattice was anisotropic with a bare
aspect ratio $a_s/a_t$ of 2. A total of 420 quenched configurations on a 
\( 10^{3}\times 30 \) lattice were analyzed. 

\begin{figure}
{\par\centering \rotatebox{0}{\includegraphics*{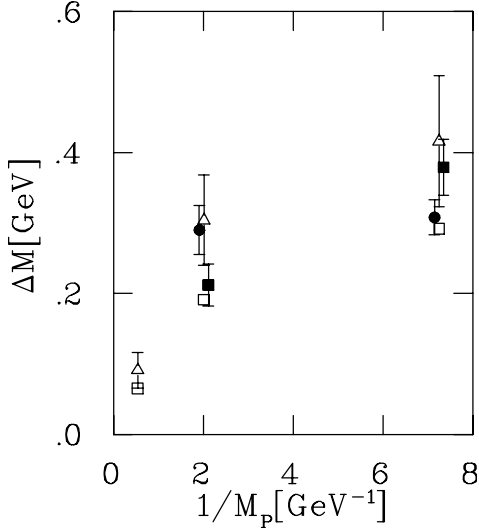}} \par}

\caption{Spin splitting for baryons with different flavour versus inverse of the
corresponding pseudoscalar meson mass. The open squares are experimental values
and the triangles are the results of this work. For comparison the results of 
\cite{cppacs} (filled squares) and \cite{ukqcd2} (filled circles) are also
shown.}
\label{fig1}
\end{figure}

The results for singly charmed
baryons are given in Table \ref{cbmasses}. In contrast to \cite{ukqcd1} the
hyperfine splittings are positive and large in magnitude. Figure \ref{fig1} shows the 
mass difference $\Sigma_c^*-\Sigma_c$ as well as the results for the 
$\Sigma$-hyperon and nucleon - delta compared to experimental values. Also
shown are the results of lattice calculations\cite{cppacs,ukqcd2} done with 
different actions on lattices with much smaller lattice spacing than used here.
The results of our simulation are seen to been quite compatible. 

\begin{table}[top]

\caption{\label{cbmasses}Masses of singly charmed baryons. Masses are given in GeV,
mass differences are in MeV. The experimental values are taken from \cite{pdg} 
except for \protect\( \Xi _{c}'\protect \) which is from \cite{jess}}
{\centering \begin{tabular}{cccc}
\hline 
&
This work&
UKQCD\cite{ukqcd1}&
Experiment\\
\hline 
\( \Lambda _{c} \)&
2.304(39)&
2.27 \( ^{+4}_{-3} \) \( _{-3}^{+3} \)&
2.285(1)\\
\( \Sigma _{c} \)&
2.465(22)&
2.46 \( ^{+7}_{-3} \) \( _{-5}^{+5} \)&
2.453(1)\\
\( \Sigma _{c}^{*} \)&
2.557(30)&
2.44 \( ^{+6}_{-4} \) \( _{-5}^{+4} \)&
2.518(2)\\
\( \Xi _{c} \)&
2.454(21)&
2.41 \( ^{+3}_{-3} \) \( _{-4}^{+4} \)&
2.468(2)\\
\( \Xi _{c}' \)&
2.579(14)&
2.51 \( ^{+6}_{-3} \) \( _{-6}^{+6} \)&
2.575(3)\\
\( \Xi _{c}^{*} \)&
2.672(16)&
2.55 \( ^{+5}_{-4} \) \( _{-5}^{+6} \)&
2.645(2)\\
\( \Omega _{c} \)&
2.664(12)&
2.68 \( ^{+5}_{-4} \) \( _{-6}^{+5} \)&
2.704(4)\\
\( \Omega _{c}^{*} \)&
2.757(14)&
2.66 \( ^{+5}_{-3} \) \( _{-7}^{+6} \)&
\\
\( \Sigma _{c}^{*}- \)\( \Sigma _{c} \)&
91(25)&
-17 \( ^{+12}_{-31} \) \( _{-2}^{+3} \)&
65(2)\\
\( \Xi _{c}^{*}-\Xi _{c}' \)&
94(13)&
-20 \( ^{+12}_{-24} \) \( _{-3}^{+2} \)&
70(4)\\
\( \Omega _{c}^{*}-\Omega _{c} \)&
94(10)&
-23 \( ^{+6}_{-14} \) \( _{-2}^{+3} \)&
\\
\hline 
\end{tabular}\par}\end{table}

The masses of doubly charmed baryons have also been calculated. The results are
3.598(13)GeV and 3.682(20)GeV for $\Xi_{cc}$ and $\Xi_{cc}^*$ respectively. For
$\Omega_{cc}$ and $\Omega_{cc}^*$ we get masses of 3.697(10)GeV and
3.775(12)GeV. The hyperfine splittings are only slightly smaller than for singly
charmed baryons.

In lattice QCD masses are extracted from correlation functions of hadron
operators. One difference between the present calculation and that of 
Ref.\cite{ukqcd1} is in the choice of baryon operators. In \cite{ukqcd1}
calculations are done with
interpolating operators of the form:
\begin{equation}
O_5=\epsilon^{abc}(l_a^TC\gamma_5l'_b)h_c, O_{\mu}=\epsilon^{abc}(l_a^TC\gamma_{\mu}l'_b)h_c
\end{equation}
where $l,l'$ are light quark fields and $h$ is a heavy field (for $O_{\mu}$,
$l'$ may be the same flavour as $l$). This choice of operators is motivated
by heavy quark symmetry where baryons are classified according to the total
spin (0 or 1) of the light flavours.

\begin{figure}
{\par \rotatebox{0}{\includegraphics*{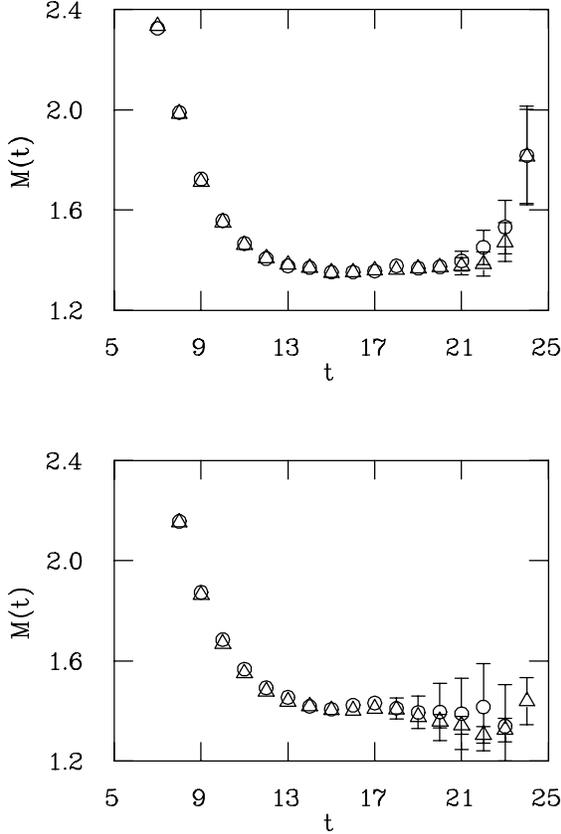}} \par}

\caption{M(t) calculated using two different baryon operators. The upper figure
is for $\Omega_c$ and the lower is for $\Omega_c^*$. The results using the
operators (1) are shown by circles, those using the operators of \cite{cbary} 
are shown by triangles.}
\label{fig2} 
\end{figure}

Our approach is different. The aim is to do a unified analysis, incorporating
all flavours (quark masses) in a common framework. We therefore start with the
nucleon operator for which a common choice\cite{leinweb1,ioffe} is
\begin{equation}
\epsilon^{abc}(u_a^TC\gamma_5d_b)u_c
\end{equation}
where $u$ and $d$ are up and down quark fields. It can be shown that this
operator contains the component
\begin{equation}
\epsilon^{abc}(u_a^TC\gamma_{\mu}u_b)\gamma_5\gamma^{\mu}d_c
\end{equation}
which is expected to survive in the limit where the mass of the unlike
quark ($d$ in this case) is taken to be very heavy. Therefore the familiar
operators used in light baryon spectroscopy should also be applicable,
with appropriate substitution of fields, in the heavy baryon sector. The
forms used may be found in \cite{cbary}.

In principle, masses should be independent of the interpolating operators
used as long as the quantum numbers are the same. In view of the difference
between the results of the present calculation and Ref.\cite{ukqcd1} it is
worthwhile to check if this independence of operator is obtained in practice.
We have therefore analyzed a subset of our configurations using operators of the
form (1) and compared the results with those obtained using the operators
given in \cite{cbary}. Typical comparisons are shown in figure \ref{fig2}.
The quantity shown in the figure is 
$M(t) = ln(G(t)/G(t+1))$ where  $G(t)$ is the hadron correlator and
for large times $t$ it is equal to the mass. It is seen that the mass is
indeed the same for the two different choices of interpolating operator. 

In the continuum limit the D234 action describes Dirac fermions. For heavy
quarks on the lattice, it can be advantageous to use a nonrelativistic
approach\cite{lnrqcd}. For b-quarks this seems to work well but for charmed 
quarks it is less clear whether lattice NRQCD is adequate\cite{split,lewis}. 
A particular concern
is that hyperfine splittings in the meson sector seem to be underestimated
compared to experimental values\cite{split,lewis,ali}. 
Therefore it is of interest to investigate
the baryon sector. 

\begin{figure}
{\par\centering \rotatebox{0}{\includegraphics*{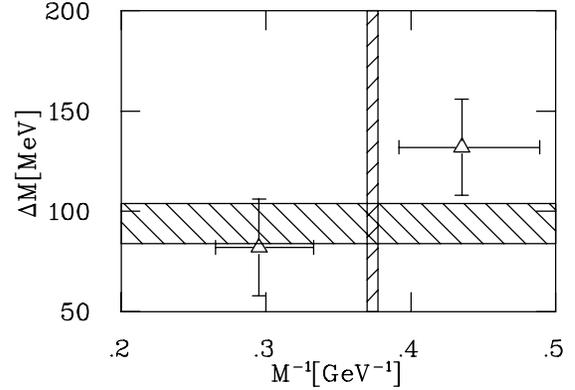}} \par}

\caption{The mass difference $\Delta M$ for $\Omega_c^*-\Omega_c$ calculated
using lattice NRQCD for the charmed quark. The intersection of the cross hatched
bands indicates the result obtained with the D234 action.}
\label{fig3} 
\end{figure}

\begin{figure}
{\par\centering \rotatebox{0}{\includegraphics*{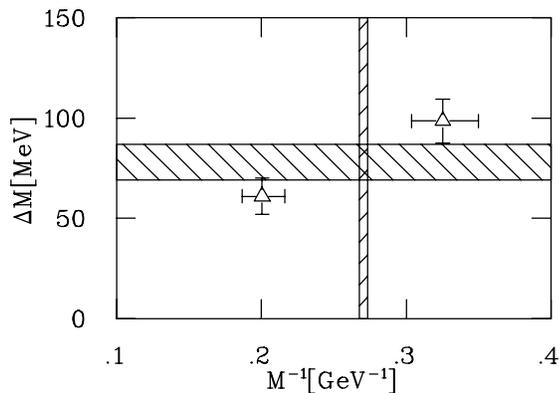}} \par}

\caption{The mass difference $\Delta M$ for $\Omega_{cc}^*-\Omega_{cc}$ 
calculated
using lattice NRQCD for the charmed quark. The intersection of the cross hatched
bands indicates the result obtained with the D234 action.}
\label{fig4} 
\end{figure}

We have carried out a preliminary analysis of charmed baryons using lattice
NRQCD, instead of the D234 action, for the charmed quark. The result for the
mass difference $\Omega_c^*-\Omega_c$ is plotted in Figure \ref{fig3} and for 
$\Omega_{cc}^*-\Omega_{cc}$ in Figure \ref{fig4}. For comparison, the results 
obtained
using the D234 action for the charmed quark are also shown. Calculations were
done for two different values of the nonrelativistic charm quark mass which
should be tuned to reproduce the charmed baryon mass (indicated by the vertical
cross hatched area in the figures). Interpolating between the two points
calculated using NRQCD, one sees that the result for the spin splitting using
NRQCD is compatible with that using the D234 ("relativistic") action.

To summarize, we have calculated the spectrum of singly and doubly charmed
baryons in quenched lattice QCD. In contrast to a previous 
calculation\cite{ukqcd1}, it was found that hyperfine splittings are positive
as expected, for example, from the quark model and are comparable in magnitude
to experimental values. This result is quite robust. It is independent of the
choice of interpolating operators for the baryons. Preliminary results also
indicate that use of an NRQCD description for the charmed quark yields values
for the spin splittings compatible with those obtained using the D234 action.

It is a pleasure to thank R. Lewis for helpful discussions. This work is
supported in part by the Natural Sciences and Engineering Research Council
of Canada.

\end{document}